\documentstyle[11pt]{article} 
\large{
\centerline{\bf Ambiguities on the quantization of a one-dimensional dissipative }
\vskip1pc
\centerline{\bf system with position depending dissipative coefficient}
}
\centerline{G. L\'opez$^{1}$,  X. E. L\'opez$^{2}$ and G. Gonz\'alez$^3$}
\centerline{$^{1}$Departamento de F\'{\i}sica de la Universidad de Guadalajara}
\centerline{Apartado Postal 4-137}
\centerline{44410 Guadalajara, Jalisco, M\'exico}   
\centerline{$^2$ Facultad de Ciencias de la UNAM}
\centerline{Apartado postal 70-348, Coyoac\'an 04511 M\'exico D.F.} 
\centerline{$^3$Departamento de Matem\'aticas y F\'{\i}sica, ITESO}
\centerline{Perif\'erico Sur 8585, C.P. 45090}
\centerline{Guadalajara, Jalisco, M\'exico}
\centerline{PACS: 03.20.+i, 03.30.+p, 03.65.-w} 
\centerline{May, 2005}
\begin{document}  
\large{
\centerline{\bf ABSTRACT}
}
\vskip1cm
For a one-dimensional dissipative system with position depending coefficient, two constant of motion are deduce.
These constants of motion bring about two Hamiltonians to describe the dynamics of same classical system. However,
their quantization describe the dynamics of two completely different quantum systems.
\vfil\eject
\large{
\leftline{\bf 1. Introduction}    
}  
\vskip0.5pc\noindent
It is well known that the Lagrangian (therefore, the Hamiltonian) formulation for some systems of more than
one dimensions may not exist [1]. Fortunately, many of our physical systems have avoided this problem and the
whole quantum and statistical mechanics of nondissipative systems are given in terms of the Hamiltonian or
Lagrangian formulation. For dissipative systems there are two main approaches, the first one consists on 
keeping
the same Hamiltonian formalism for the system and the interacting background. By doing this, one brings about a
master equation, and the dissipation and diffusion parameters appear as a part of the solution [2]. This approach
has its own merits, but it will not be fallowed on this paper. We will follow the second approach which
consists in
obtaining a phenomenological velocity depending Hamiltonian which represents the classical dissipative system 
and proceeding to make the usual quantization with this Hamiltonian. Within this approach, one can additionally
study  the mathematical consistence of the Hamiltonian quantum mechanics. Now, even for
one-dimensional systems, where the existence of their Lagrangians is guaranteed [3], the Lagrangian and
Hamiltonian formulation are not free of problems [4]. One of the problems we are concerned is the
implication on the quantization of a classical system when different Hamiltonians describe the same
classical system [5]. This problem would represent an ambiguity on the quantum mechanics for the quantization
of a classical system. To study this problem, we consider a one-dimensional velocity depending force which
may represent a dissipative system with proper selection of parameters. Following the procedure given in
reference [6], two constant of motion with energy units are found for this system. Their associated Lagrangians
and Hamiltonians are deduced, and, using the perturbative quantum theory, the two resulting  quantum
dynamics are shown with these Hamiltonian.
\vskip2pc
\leftline{\bf 2. Constants of Motion}
\vskip1pc\noindent
Let us consider a one-dimensional system described by the following equation
$$m{d^2x\over dt^2}=-U_x(1-\alpha {\dot x}^2)\ ,\eqno(1)$$
where $m$ is the mass of the particle, $x$ and $\dot x$ denote its position and velocity, $U_x$ represents the
differentiation with respect to $x$ of the function $U(x)$, and $\alpha$ is a real parameter. This equation
could represent the motion of a particle under a force $-U_x$ in a dissipative medium characterized by a
quadratic velocity force (for $\dot x>0$ and $\alpha<0$) with a coefficient depending on the position of the
form $\alpha U_x$. Eq. (1) can be written as the following autonomous dynamical system
$${dx\over dt}=v\hskip2pc {dv\over dt}=-U_x(1-\alpha v^2)/m\eqno(2)$$
which has the set $S_c=\{(x_c,0)\in\Re^2|~U_x(x_c)=0\}$ as the set of critical points in the phase space
$(x,v)$~[7], which is the same as that for $\alpha=0$.  A constant of motion for this system is a function
$K=K(x,v)$ such that
$dK/dt=0$, i.e., it satisfies the following partial differential equation [8]
$$v{\partial K\over\partial x}-{U_x\over m}(1-\alpha v^2){\partial K\over\partial v}=0\ .\eqno(3)$$
The general solution of this equation is given by [9]
$$K_{\alpha}(x,v)=G(C)\ ,\eqno(4)$$
where $G$ is an arbitrary function of the characteristic curve of (3), denoted by $C$. This characteristic
curve can by deduced as 
$$C=-{m\over 2\alpha}\log(1-\alpha v^2)+U(x)\ .\eqno(5)$$
The function  $G$ can be chosen such that in the limit for the dissipation parameter goes to zero, one can get
the usual energy expression of a conservative system,
$E=\lim_{\alpha\to 0}K_{\alpha}=mv^2/2+U(x)$. There are at least two ways to do this. One way is by selecting $G$
as the identity function, $G(C)=C$, and the other is by selecting $G$ of the form $G(C)=e^{-2\alpha
C/m}/\alpha+m/2\alpha$. These two expressions for $G$ bring about the following constants of motion
$$K_{\alpha}^{(1)}(x,v)= -{m\over 2\alpha}\log(1-\alpha v^2)+U(x)\eqno(6)$$
and
$$K_{\alpha}^{(2)}(x,v)={1\over 2}mv^2e^{-2\alpha U(x)/m}+{m\over 2\alpha}\biggl(1-e^{-{2\alpha U(x)/m}}\biggr)\
.\eqno(7)$$
\vskip2pc
\leftline{\bf 3. Lagrangian, Generalized Linear Momentum and Hamiltonian}
\vskip1pc\noindent
Using the known expression [10] to get the Lagrangian given the constant of motion,
$$L(x,v)=v\int{K(x,v)~dv\over v^2}\ ,\eqno(8)$$
the Lagrangians associated to (6) and (7) are
$$L_{\alpha}^{(1)}(x,v)={mv\over\sqrt{\alpha}}arctanh(v\sqrt{\alpha})+
{m\over 2\alpha}\log(1-\alpha v^2)-U(x)\eqno(9)$$
and 
$$L_{\alpha}^{(2)}(x,v)={1\over 2}mv^2e^{-{2\alpha U(x)/m}}-{m\over 2\alpha}\biggl(1-e^{-{2\alpha U(x)/m}}
\biggr)\ .\eqno(10)$$
Their generalized linear momenta ($p=\partial L/\partial v$) are
$$p_{\alpha}^{(1)}={m\over\sqrt{\alpha}}arctanh(v\sqrt{\alpha})\eqno(11)$$
and
$$p_{\alpha}^{(2)}=mv~e^{-2\alpha U(x)/m}\ .\eqno(12)$$
Thus, their associated Hamiltonians, 
$H_{\alpha}^{(1)}(x,p_{\alpha}^{(i)})=K_{\alpha}^{(i)}\bigl(x,v(x,p_{\alpha}^{(i)})\bigr)$, for \break $i=1,2$  
are given by
$$H_{\alpha}^{(1)}(x,p)=-{m\over 2\alpha}\log\left(1-\tanh^2\left({p\sqrt{\alpha}\over m}\right)\right)+U(x)
\eqno(13)$$
and
$$H_{\alpha}^{(2)}(x,p)={p^2\over 2m}~e^{2\alpha U(x)/m}+{m\over 2\alpha}\left(1-e^{-2\alpha U(x)/m}\right)\
,\eqno(14)$$
where the same $p$ variable has been used for the generalized linear momentum to simplify the notation.
Note from (9), (10), (11), (12), (13) and (14) that the following limits are satisfied
$$\lim_{\alpha\to 0}L_{\alpha}^{(i)}(x,v)={1\over 2}mv^2-U(x)\ ,\ i=1,2\eqno(15a)$$
$$\lim_{\alpha\to 0}p_{\alpha}^{(i)}=mv\ ,\ i=1,2\eqno(15b)$$
and
$$\lim_{\alpha\to 0}H_{\alpha}^{(i)}(x,p)={p^2\over 2m}+U(x)\ ,\ i=1,2\ ,\eqno(15c)$$
corresponding to the conservative case of (1). At first order on the parameter $\alpha$, one gets from (13) and
(14)  the following Hamiltonians
$$H_1={p^2\over 2m}+U(x)+\alpha{19 p^4\over 48 m^3}\eqno(16)$$
and
$$H_2={p^2\over 2m}+U(x)+\alpha\left[{p^2U(x)\over m^2}-{2\over m}U^2(x)\right]\ .\eqno(17)$$
We must point out that the Lagrangians (9) and (10), the generalized momenta (11) and (12), and the
Hamiltonians (13) and (14) are associated to the same system (1), that is, they generate the same classical
dynamics.
\vskip2pc
\leftline{\bf 4. Quantization}
\vskip1pc\noindent
It is clear that Sch\"odinger quantization of the system (1) using the corresponding Hermitian operator
to the Hamiltonians (13) or (14),
$$i\hbar{\partial \Psi\over\partial t}=\hat H_{\alpha}^{(i)}\Psi\ ,\eqno(18)$$
will generate different quantum dynamics. Therefore, it is just necessary to show this difference at first
order in the parameter $\alpha$ with the Hamiltonians (16) and (17) and at first order in perturbation theory
[11]. Since (16) and (17) are time independent, one uses time independent perturbation theory. According to this
theory, the corrections to the nth-eigenvalue and nth-eigenvector  are given by
$$E_n^{(i)}=E_n^{(0)}+\langle n|\widehat W_i|n\rangle\ ,i=1,2\eqno(19a)$$
and
$$\psi_n^{(i)}(x)=\psi_n^{(0)}(x)+\sum_{k\not=n}{\langle n|\widehat W_i|k\rangle\over
E_n^{(0)}-E_k^{(0)}}~\psi_k^{(0)}(x)\ ,i=1,2,\eqno(19b)$$
where the Hermitian operators associated to the  Hamiltonians (16) and (17) are written as $$\widehat
H_i=\widehat H_0+\widehat W_i\ ,  i=1,2\eqno(20)$$ with $\widehat H_0$ and $\widehat W_i$ given by
$$\widehat H_0={{\hat p}^2\over 2m}+U(x)\ ,\eqno(21a)$$
$$\widehat W_1=\alpha{19{\hat p}^4\over 48 m^3}\ ,\eqno(21b)$$
and
$$\widehat W_2=\alpha\left[{{\widehat {p^2U(x)}}\over m^2}-{2\over m} U^2(x)\right]\ .\eqno(21c)$$
The energy $E_n^{(0)}$ and the wave function $\psi_n^{(0)}(x)$ are solutions of the eigenvalue problem
$$\widehat H_0\psi_n^{(0)}(x)=E_n^{(0)}\psi_n^{(0)}(x)\ .\eqno(22)$$
One can also use the Dirac notation $\widehat H_0|n\rangle=E_n^{(0)}|n\rangle$, as it has been used in (19a) and
(19b). 
\vskip1pc\noindent
To get explicitly the values for (19a) and (19b), let us consider the harmonic oscillator potential
$U(x)=m\omega^2 x^2/m$, where $\omega $ is its natural frequency of oscillation. For this case, one can write
(20) in terms of ascent and descent operators, $a^+$ and $a$, as 
$$\widehat H_0=\hbar\omega\left(a^+a+{1\over 2}\right)\ ,\eqno(23)$$
where the operators $a^+$ and $a$ have been defined as
$$a^+=\sqrt{m\omega\over 2\hbar}~x-{i\over\sqrt{2m\hbar\omega}}\hat p\eqno(24a)$$
and
$$a=\sqrt{m\omega\over 2\hbar}~x+{i\over\sqrt{2m\hbar\omega}}\hat p\eqno(24b)$$
which satisfy the following relations
$$[a^+,a]=1,\hskip0.5pc a^+|n\rangle=\sqrt{n+1}~|n+1\rangle,\hskip0.5pc a|n\rangle=\sqrt{n}~|n-1\rangle,
\hskip0.5pc a|0\rangle=0\ .\eqno(25a)$$
In addition, $E_n^{(0)}$ is given by
$$E_n^{(0)}=\hbar\omega\left(n+{1\over 2}\right)\ .\eqno(25b)$$
Using Weyl idea, one has that the Hermitian operator associated to the product $p^2U(x)$ can be given by
$$\widehat{ {p^2U(x)} }={m\omega^2\over 12}\bigl[x^2\hat p^2+\hat p^2 x^2+x\hat p x\hat p+\hat p x \hat p x+ x\hat
p^2 x+\hat p x^2\hat p\bigr]\ .\eqno(26)$$
Therefore, written $x$ and $\hat p$ in terms of $a^+$ and $a$,
$$x=\sqrt{\hbar\over 2m\omega}~(a^++a)\ ,\hskip1pc \hat p=i\sqrt{m\hbar\omega\over 2}~(a^+-a)\ ,\eqno(27)$$
and using (25a), one has
$$\langle n|\widehat W_1|n\rangle=\alpha{19\hbar^2\omega^2\over 192 m}(4n^2+4n+3)\eqno(28a)$$
and
$$\langle n|\widehat W_2|n\rangle=-\alpha{\hbar^2\omega^2\over 4m}(2n^2+2n+1)\ .\eqno(28b)$$
Thus, the modifications to the eigenvalues at first order in perturbation theory are
$$E_n^{(1)}=\hbar\omega\left(n+{1\over 2}\right) +\alpha{19\hbar^2\omega^2\over 192 m}(4n^2+4n+3)\eqno(29a)$$
and
$$E_n^{(2)}=\hbar\omega\left(n+{1\over 2}\right)-\alpha{\hbar^2\omega^2\over 4m}(2n^2+2n+1)\ .\eqno(29b)$$
Similarly, the modifications to the eigenvectors are
$$\psi_n^{(1)}(x)=|n\rangle+\alpha{19\hbar\omega\over 768 m}\biggl[A_{-4}|n-4\rangle-2A_{-2}|n-2\rangle-
2A_2|n+2\rangle-A_4|n+4\rangle\biggr]\eqno(30a)$$
and
$$\psi_n^{(2)}(x)=|n\rangle-\alpha{\hbar\omega\over 16 m}\biggl[A_{-4}|n-4\rangle+A_{-2}|n-2\rangle+
A_2|n+2\rangle+A_4|n+4\rangle\biggr]\ ,\eqno(30b)$$
where $A's$ have been defined as

\begin{eqnarray*}
A_{-4}&=&\sqrt{n(n-1)(n-2)(n-3)}\ ,\\
A_{-2}&=&(n-2)\sqrt{(n-1)(n+1)}+(n-1)\sqrt{n(n-1)}+(2n-1)\sqrt{(n-2)(n-1)}\ ,\\
A_2&=&(n+1)\sqrt{n(n+2)}+(n+3)\sqrt{(n+1)(n+2)}+(2n+6)\sqrt{n+2)(n+1)}\ ,\\
A_4&=&\sqrt{(n+4)(n+3)(n+2)(n+1)}\ ,\\
\end{eqnarray*}
As one can see from (29) and (30), the quantum dynamics of the system (13) and (14) are quite different even a
first order in perturbation theory and at first order in the parameter $\alpha$.
\vskip2pc
\leftline{\bf 5. Conclusions and comments}
\vskip1pc\noindent
For a one-dimensional dissipative system with position depending coefficient, we have shown that it is possible
to have two  different Hamiltonians associated to the same classical dynamics, but they bring about
completely different quantum dynamics. Since this happens for any potential $U(x)$ and due to the
expression (13) and (14), this ambiguity occurs for a non numerable set of one-dimensional systems. In addition,
one has implication on cahotic systems, since the classical equation of motion for the cahotic systems [12] 
may be formulated can be formulated using different nonequivalent time depending Hamiltonians (for
one-dimensional cahotic motion), the behavior  of the corresponding quantum system would have different types of
quantum chaos.  Finally, one needs to point out here that this ambiguity will appear also on the Hamiltonian
formulation of classical and quantum statistical mechanics.
\vskip2pc
\leftline{\bf Acknowledgments}
\vskip1pc\noindent
One of us, G. L\'opez, wants to thank V.I. Man'ko for his very  useful discussions. 
\vfil\eject
\leftline{\bf References}
\vskip2pc\noindent
\obeylines{
1. J. Douglas, Trans. Amer. Math. Soc.,{\bf 50} (1941) 71.
2. A. O. Caldeira and A.T. Legget, Physica A, {\bf 121} (1983) 587.
\quad W.G. Unruh and W.H. Zurek, Phys. Rev. D, {\bf 40} (1989) 1071.
\quad B.L. Hu, J.P. Paz and Y. Zhang, Phys. Rev. D, {\bf 45} (1992) 2843.
3. D. Darboux,{\it Le\c cons sur la th\'eorie g\'en\'eral des surfaces et les applications 
\quad g\'eom\'etriques du calcul infinit\'esimal}, IVi\'eme partie,
\quad Gauthoer-VIllars, Paris, 1984.
4. P. Havas, Act. Phys. Austr., {\bf 38} (1973) 145.
\quad S. Okubo, Phys. Rev. D {\bf 22} (1980) 919.
\quad V.V. Dodonov, V.I. Man'ko and V.D. Skarzhinsky,
\quad Hadr. Jour. {\bf 4} (1981) 1734.
\quad G. Marmon, E.J. Salentan, A. Simoni and B. Vitale, {\it Dynamical System:
\quad a Differential Geometric Approach}, Wiley, Chinchester, 1985.
\quad R. Glauber and V.I. Man'ko, Sov. Phys. JEPT, {\bf 60} (1984) 450.
\quad G. L\'opez, Intr. Jour. Theo. Phys.,{\bf 37},5 (1998) 1617.
\quad G. L\'opez, Rev. Mex. Fis., {]bf 45},6 (1999) 551.
5. G. L\'opez, Rev. Mex. Fis., {\bf 48},1 (2002) 10.
6. G. L\'opez, Ann. Phys., {\bf 251}, 2(1996) 372. 
7. P.G. Drazin, {\it Nonlinear Dynamics}, Cambridge University Press, 1992.
8. G. L\'opez, {\it Partial Differential Equations of First Order and Their
\quad Applications to Physics}, World Scientific, 1999.
9. F. John, {\it Partial Differential Equations}, Springer-Verlag, N.Y. 1974.
10. J.A. Kobussen, Act. Phys. Austr., {\bf 51} (1979) 193.
\quad C. Leubner, Phys. Rev. A {\bf 86} (1987) 9.
\quad C.C. Yan, Amer. J. Phys., {\bf 49} (1981) 296.
\quad G. L\'opez, Ann. Phys., {\bf 251}, 2(1996) 372.
11. A. Messiah, {\it Quantum Mechanics Vol. II}, John-Wiley and Sons, 1958.
12. G. Casati and B. Chirikov (Eds.),{\it Quantum Chaos: between order 
\quad and disorder}, Cambridge University Press, 1995.
\quad F. Benvenutto and G. Casati, Phys. Rev. Lett., {\bf 72} (1994) 1818.
\quad E. Ott, {\it Chaos in Dynamical Systems}, Cambridge University Press, 1993.
\quad G.M. Zaslavsky and N.N. Filonenko, Zh., Eksp. Teor. 
\quad Fiz. {\bf 65} (1973) 643.
\quad K.N. Alekseev et al, Sov. Phys. Uspekhi, {\bf 35] (1992) 572.
\quad R. Graham and M. H\"ohnerbach, Phys. Lett. A, 
\quad {\bf 101},2 (1984) 61. 
}

\end{document}